\documentclass[iop]{emulateapj}
\usepackage{apjfonts}
\usepackage[breaklinks,colorlinks,urlcolor=black,citecolor=black,linkcolor=black]{hyperref}
 \usepackage{alltt}

\usepackage{hyperref}
\usepackage{graphics,graphicx,url,verbatim,xspace}
\usepackage{bold-extra}
\usepackage[usenames, dvipsnames]{color}
\usepackage{enumitem}
\usepackage{natbib}
\usepackage{lipsum}
\usepackage{eurosym}
\usepackage{fancyhdr}
\usepackage{multirow}
\usepackage{color}
\usepackage{colortbl}
\newcommand\G{\rowcolor[gray]{0.9}[\tabcolsep][\tabcolsep]}

\definecolor{black}{rgb}{0,0,0}
\definecolor{red}{rgb}{1.0,0,0}

\newcommand{\specialcell}[2][c]{%
  \begin{tabular}[#1]{@{}c@{}}#2\end{tabular}}

\setlist[itemize]{noitemsep, topsep=0pt}
\newcommand\T{\rule{0pt}{3.1ex}}       
\newcommand\B{\rule[-1.7ex]{0pt}{0pt}} 

\newcommand{\BLI}{\textit{Breakthrough Listen Initiative}\xspace}
\newcommand{\BL}{\textit{Breakthrough Listen}\xspace}

\newcommand{\ft}[1]{ #1}

\definecolor{black}{rgb}{0,0,0}
\definecolor{red}{rgb}{1.0,0,0}

\begin{document}
\pagestyle{plain}
\pagenumbering{arabic}


\author{J. Emilio Enriquez \altaffilmark{1,2},
Andrew Siemion\altaffilmark{1,2,3},
Griffin Foster\altaffilmark{1,6},
Vishal Gajjar\altaffilmark{4},
Greg~Hellbourg\altaffilmark{1},\\
Jack Hickish\altaffilmark{5},
Howard Isaacson\altaffilmark{1},
Danny C. Price\altaffilmark{1,7},
Steve Croft\altaffilmark{1},
David DeBoer\altaffilmark{5},\\
Matt~Lebofsky\altaffilmark{1},
David H. E. MacMahon\altaffilmark{5},
Dan Werthimer\altaffilmark{1,4,5}
}

\altaffiltext{1}{Department of Astronomy, University of California, \\ Berkeley, 501 Campbell Hall \#3411, Berkeley, CA, 94720, USA \\ e.enriquez@berkeley.edu}
\altaffiltext{2}{Department of Astrophysics/IMAPP, Radboud University, P.O. Box 9010, NL-6500 GL Nijmegen, The Netherlands}
\altaffiltext{3}{SETI Institute, Mountain View, CA 94043, USA}
\altaffiltext{4}{Space Science Laboratory, 7-Gauss way, University of California, Berkeley, CA, 94720, USA}
\altaffiltext{5}{Radio Astronomy Laboratory, University of California at Berkeley, Berkeley, CA 94720, USA}
\altaffiltext{6}{University of Oxford, Sub-Department of Astrophysics, Denys Wilkinson Building, Keble Road, Oxford, OX1 3RH, United Kingdom}
\altaffiltext{7}{Centre for Astrophysics \& Supercomputing, Swinburne University of Technology, PO Box 218, Hawthorn, VIC 3122, Australia}

\title{The Breakthrough Listen Search for Intelligent Life: \\1.1-1.9 GHz observations of 692 Nearby Stars}


\begin{abstract}
We report on a search for engineered signals from a sample of 692 nearby stars using the Robert C.\ Byrd Green Bank Telescope, undertaken as part of the \BLI search for extraterrestrial intelligence. Observations were made over 1.1$-$1.9 GHz ($L$ band), with three sets of five-minute observations of the 692 primary targets, interspersed with five-minute observations of secondary targets. By comparing the ``ON'' and ``OFF'' observations we are able to identify terrestrial interference and place limits on the presence of engineered signals from putative extraterrestrial civilizations inhabiting the environs of the target stars. During the analysis, eleven events passed our thresholding algorithm,  but a detailed analysis of their properties indicates they are consistent with known examples of anthropogenic radio frequency interference.  We conclude that, at the time of our observations, none of the observed systems host high-duty-cycle radio transmitters emitting between 1.1 and 1.9 GHz with an Equivalent Isotropic Radiated Power of $\sim10^{13}$ W, which is readily achievable by our own civilization.  Our results suggest that fewer than $\sim$ 0.1$\%$ of the stellar systems within 50\,pc possess the type of transmitters searched in this survey. 

\end{abstract}
\maketitle

\section{Introduction.}
\label{sec:intro}

The question of whether or not the Earth is alone in the universe as a host for life is among the most profound questions in astronomy. The question's profundity occupies a singular place in any conception of the human relation with the cosmos. The search for life and Earth-like environments has long received a great deal of attention from astronomers, punctuated  most recently by  a series of  discoveries that have determined conclusively that Earth-like exoplanets exist in abundance throughout our galaxy \citep{2013ApJ...767...95D,2013PNAS..11019273P,2014PNAS..11112647B}.

The search for life beyond the Earth, either extinct or extant, is currently pursued via three primary means: direct in-situ detection of life or the byproducts of biological processes in nearby environments (e.g. the subsurface of Mars, \citealt{2015Sci...347..415W}); remote sensing of biological activity in gaseous plumes from nearby bodies \citep{2014Sci...343..171R}, exoplanet atmospheres and surfaces \citep{2014PNAS..11112634S}; or by detecting --- either directly or indirectly --- the presence of technology produced by an extraterrestrial intelligence \citep{Tarter:2003p266}.

In situ searches for life signatures, while naturally allowing an incredible range of possible investigations, are severely limited in their range from an astronomical perspective.  Even the most ambitious planned in-situ astrobiology missions could only hope to reach the nearest few stars to Earth and would take several dozen years to do so. Remote spectroscopic sensing of the atmospheres of Earth-like exoplanets offers more immediate opportunities, but the extreme difficulty of attaining a sufficiently significant detection of potentially biotic constituents limits this technique to a handful of potential targets out to perhaps 10~pc \citep{2005AsBio...5..706S,2014ApJ...781...54R,2016ApJ...819L..13S}. Even for those targets amenable to remote spectroscopic searches for biology, necessary exposure durations with next-generation telescopes (e.g. The \textit{James Webb Space Telescope} (JWST\footnote{\url{https://www.jwst.nasa.gov}}), Thirty Meter Telescope (TMT\footnote{\url{http://www.tmt.org}}), Giant Magellan Telescope (GMT\footnote{\url{http://www.gmto.org}}) and the European Extremely Large Telescope (E$-$ELT\footnote{\url{http://www.eso.org/sci/facilities/eelt/}})) are measured in days, and detections potentially suffer from confusion with abiotic processes that may give rise to similar signatures. 

Searches for intelligent life targeting signatures of technology are unique in their ability to probe the entire observable universe given appropriate assumptions about the transmitting technology. Importantly, the generation of extremely luminous emission, detectable over a large portion of our galaxy with humanity's observing capabilities, is possible using zero or minimal extrapolation from humanity's current technological capacity.

\cite{1961PhT....14...40D} and others have developed frameworks to estimate how many civilizations exist in the galaxy. However, given the current uncertainties, it is equally likely that there are thousands of civilizations in the galaxy or that we are the only one. Only the covered sample size of large surveys can shed light on this question.

Such surveys are technologically and logistically challenging. These challenges arise from the unknown frequency distribution, duty cycle, and luminosity function of putative transmissions. The potential spectral similarity between anthropogenic radio-frequency interference (RFI) and extraterrestrial technological transmissions brings additional complications. The sheer immensity of the parameter space that must be explored is a potential explanation to the absence of radio detections of extraterrestrial intelligence, despite numerous previous efforts 
\citep{1973Icar...19..329V,1980Icar...42..136T,1983Icar...53..147B,1986Icar...67..525H,1994Icar..107..215S,1995ASPC...74..285B,1996A&A...306..141M,1998AcAau..42..651B,2000ASPC..213..479W,2011arXiv1109.1595K,2013ApJ...767...94S,2016AJ....152..181H,2017AJ....153..110G}.

Early radio SETI experiments used only a narrow frequency band relative to modern wide-band radio telescope observing systems. This influenced those efforts into concentrating searches at or near specific frequencies of interest. The most common examples are the searches around known energy transitions such as the hydrogen hyperfine transition line at 21 cm \citep{1959Natur.184..844C}, the hydroxyl lines around 18 cm \citep{1980Icar...42..136T}, the spin-flip line frequency of positronium \citep{1994Icar..107..215S,1996A&A...306..141M}, and the tritium hyperfine line \citep{1986Icar...65..152V}. ``Magic'' frequencies around numerical combinations of special cosmological constants have also been proposed \citep{1973Natur.245..257D}.  Progress in radio instrumentation allows modern radio telescopes to survey much wider frequency bandwidths over much larger areas of sky for a fixed observation time. This has the potential to significantly reduce the inherent bias in selecting specific regions of the radio spectrum.

The \BLI, announced in 2015, uses the Automated Planet Finder optical telescope as well as two radio telescopes --- the Parkes Telescope in Australia and the Robert C. Byrd Green Bank Telescope (GBT) in West Virginia --- to scan the sky for technosignatures. \BL aims to survey one million stars selected from several sub-samples, including several thousands of the nearest stars to the Sun selected for detailed study with all three facilities \citep{2017PASP..129e4501I}.  The \BLI has more recently also announced partnerships with two additional facilities, the FAST 500m telescope under construction in Southern China\footnote{\url{https://breakthroughinitiatives.org/News/6}} and the Jodrell Bank Observatory and University of Manchester in the United Kingdom.\footnote{\url{https://breakthroughinitiatives.org/News/11}}

In this paper, we report the first search for engineered signals of extraterrestrial origin using data from the \BL project. This work represents the first of a series of data and detection releases for the \BL project. The data and analysis pipelines used in the \BL project are open access, and we aim to provide a regular update on the ongoing surveys and analysis techniques. All data and observational information used in this work can be found at the survey website\footnote{\url{http://seti.berkeley.edu/lband2017/}}. The paper is organized as follows. In Sec.~\ref{sec:obs} we present the observational strategy and provide a brief overview of digital hardware. Data analysis techniques and algorithms are discussed in Sec. \ref{sec:analysis}; results are presented in Sec.~\ref{sec:results}. In Sec.~\ref{sec:discussion} we discuss the results in the context of previous SETI efforts.
The paper concludes with a summary of the results and limits one may place upon narrowband transmissions based on this work, before giving a summary of future plans and closing remarks.\\

\vspace{30px}
\section{Observations}
\label{sec:obs}

A sample of nearby stellar targets for this campaign was selected from the \textit{Hipparcos} catalog
 \citep{1997A&A...323L..49P}. The total number\footnote{\cite{2017PASP..129e4501I} published number is 1709, but 1702 is the total number after removing double counting from some binary stars.} of target stars described in \cite{2017PASP..129e4501I} is 1702. Of these, 1185 are observable with the GBT. A subset of 692, for which we have good quality data and a full cadence is analyzed in this paper, a representative list is presented in Table \ref{table:obs_table}.\footnote{The full table is available in the online version of this article.} 

The observations of the sample of stars presented in this work were taken between January 2016 and February 2017 with the L-band receiver at GBT, covering between 1.1 and 1.9 GHz. We employed the available notch filter between 1.2 and 1.33 GHz to exclude strong local radar signals. The range of frequencies of the L-band receiver covers the entire "water hole" \citep[1.4--1.7 GHz;][]{1971asee.nasa.....O}. This region, well known in the SETI literature, is bounded by the hydrogen hyperfine transition line near 21~cm ($\sim$1420 MHz) at the lower end and the four hydroxyl lines near 18~cm ($\sim$1700 MHz) at the higher end.

\begin{table*}
\centering
\caption{Truncated table of the 692 star systems observed at 1.1$-$1.9 GHz (L band) with GBT in This Work.}

\begin{tabular}{lcccccl}
\hline
\B\T Star Name & R.A. [J2000] & Decl. [J2000] & Spectral Type & Distance (pc) & UT Date & Off-source Targets \\
\hline
\hline
\G HIP 113357&  22:57:28.2  &  +20:46:08.0  &G5V&15.30&2016 Jan 02&\verb|HIP113357_OFF, HIP113357_OFF, HIP113357_OFF|\\
HIP 113368&  22:57:39.5  &  -29:37:22.1  &A3V&7.68&2016 Jan 13&\verb|HIP113368_OFF, HIP113368_OFF, HIP113368_OFF|\\
\G HIP 2422&  00:30:56.7  &  +77:01:08.8  &K0IV&39.40&2016 Jan 14&\verb|HIP2422_OFF, HIP2422_OFF, HIP2422_OFF|\\
HIP 2552&  00:32:34.2  &  +67:14:03.8  &M2.5Ve&10.10&2016 Jan 14&\verb|HIP2552_OFF, HIP2552_OFF, HIP2552_OFF|\\
\G HIP 11048&  02:22:15.0  &  +47:52:48.4  &M2&11.90&2016 Jan 16&\verb|HIP11048_OFF, HIP11048_OFF, HIP11048_OFF|\\
HIP 11090&  02:22:50.0  &  +41:23:45.2  &F0III-I&47.20&2016 Jan 16&\verb|HIP11090_OFF, HIP11090_OFF, HIP11090_OFF|\\
\G HIP 32769&  06:49:57.5  &  +60:20:14.6  &M0p&16.40&2016 Jan 16&\verb|HIP32769_OFF, HIP32769_OFF, HIP32769_OFF|\\
HIP 32919&  06:51:31.9  &  +47:21:53.3  &K2&18.80&2016 Jan 16&\verb|HIP32919_OFF, HIP32919_OFF, |\\
\G ... & & & & & & \\
HIP 114622&  23:13:20.8  &  +57:10:11.3  &K3Vvar&6.52&2017 Feb 19&\verb|HIP113764, HIP113716, HIP113755|\\
\G HIP 1086&  00:13:30.5  &  +41:02:03.5  &F0IV&35.00&2017 Feb 19&\verb|HIP1125, HIP1152, HIP1233|\\
HIP 1368&  00:17:06.8  &  +40:56:55.0  &M0&14.90&2017 Feb 19&\verb|HIP1125, HIP1152, HIP1233|\\
\G HIP 3206&  00:40:49.4  &  +40:11:04.2  &K2V&17.20&2017 Feb 19&\verb|HIP2258, HIP2420, HIP2434|\\
HIP 428&  00:05:12.3  &  +45:47:05.6  &M2&11.40&2017 Feb 19&\verb|HIP1090, HIP1337, HIP1343|\\
\G HIP 4436&  00:56:45.2  &  +38:29:55.7  &A5V&41.70&2017 Feb 19&\verb|HIP3333, HIP3597, HIP3677|\\
HIP 4907&  01:02:58.3  &  +69:13:34.0  &G5&25.80&2017 Feb 20&\verb|HIP3876, HIP4550, HIP4635|\\
\G HIP 97222&  19:45:33.6  &  +33:35:59.6  &K3V&20.30&2017 Feb 20&\verb|HIP98126, HIP97744, HIP97891|\\
GJ 725&  18:42:44.0  &  +59:38:01.7  &M3.0V&3.52&2017 Feb 20&\verb|HIP91052, HIP91065, HIP91136|\\
\hline
\end{tabular}
\label{table:obs_table}
\textbf{Note.}
Stars are identified by either the Hipparcos catalog id (prefixed with ``HIP'') or the Gliese\---Jahrei\ss \,catalog id (prefixed with ``GJ''), along with the (R.A., Decl.) sky position, spectral type, and distance from the Earth in parsecs. The last two columns list the UT date of the observation with the GBT, and the off-source targets observed using our on/off observation strategy, see Section \ref{sec:strategy} for details. A full list can be found at the survey website: \url{http://seti.berkeley.edu/lband2017}/
\end{table*}

\subsection{Strategy}
\label{sec:strategy}
Our current targeted observing strategy for the GBT and Parkes, and that employed for the analysis described here, consists of three five-minute observations of each target drawn from a primary sample set \citep{2017PASP..129e4501I}, interspersed with five-minute observations of one or more locations at least six beamwidths away from the primary source, which is beyond the primary and side lobes of the GBT and Parkes beams. 
Artificial signals that are only present in the three observations of a given primary target (i.e. the \textit{``\texttt{ON}''} observations), but are absent in the \textit{``\texttt{OFF}''} observations, are less likely to be RFI compared to signals, which are expected to affect both \textit{``\texttt{ON}''} and \textit{``\texttt{OFF}''} sources similarly if arising from emission detected in the side lobes of the beam.

\begin{figure}[htb]
\begin{center}
    \begin{tabular}{c}
        \includegraphics[width=1.0\linewidth]
{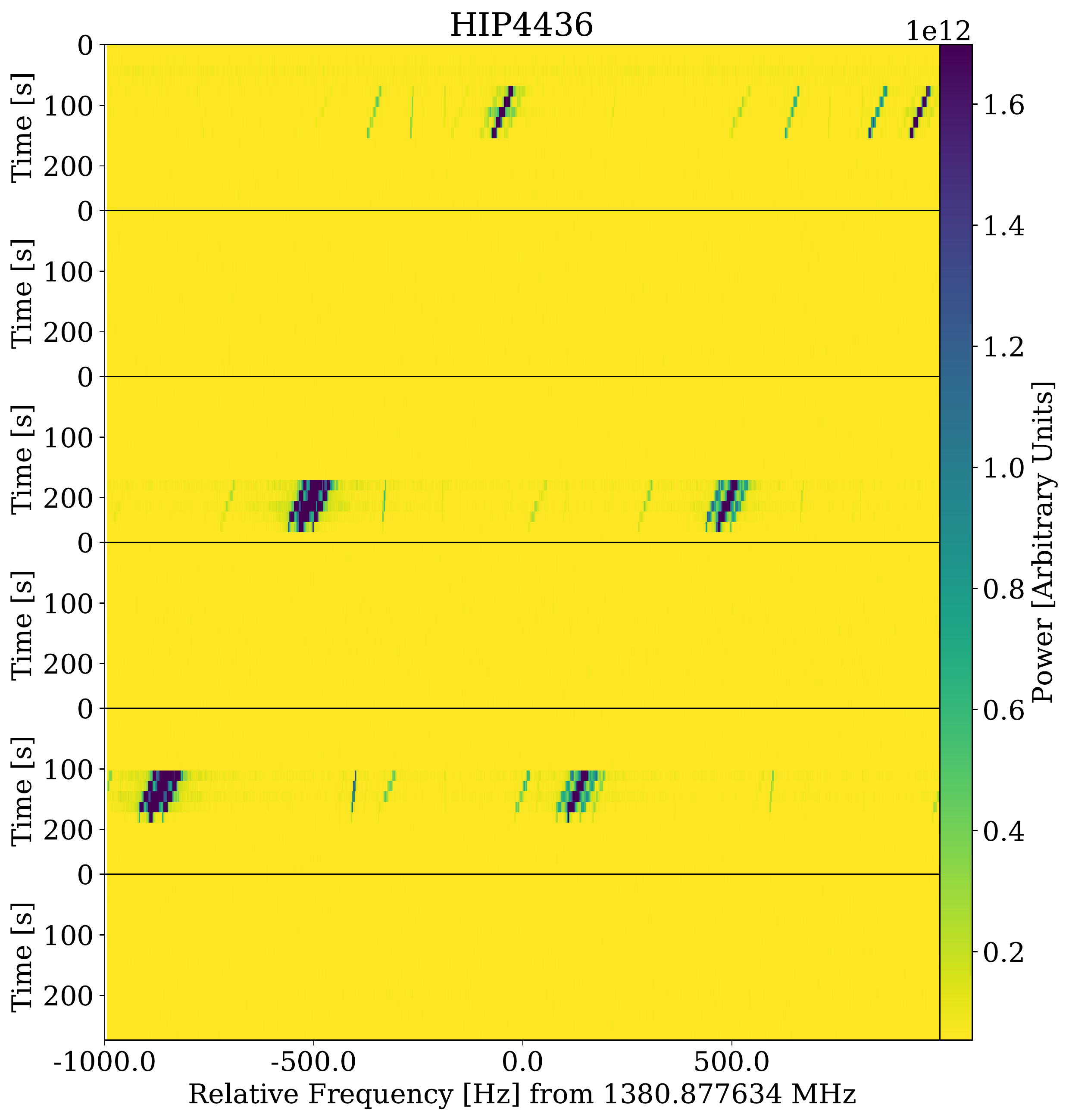}
\end{tabular}
    \caption{\footnotesize{Highlight of a detected signal over a series of 3 $\times$ 5 minute ABACAD observations of HIP 4436. The \textit{"\texttt{OFF}"} observations targeted HIP 3333, HIP 3597, and HIP 3677. Figures \ref{fig:waterfall_HIP65352} and \ref{fig:waterfall_HIP7981} below show other above-threshold events,  the observations follow the ABACAD strategy. }}
    \label{fig:waterfall_HIP4436}
\end{center}
\end{figure}

Two observation strategies were adopted. The first strategy required that on-source targets were interspersed with off-source pointings at a constant offset in declination from the primary source. This approach is referred to as ABABAB. In order to have better coverage of any potential sidelobe effects, we developed a second strategy that consisted of having the off-source targets drawn from a secondary sample list of the \textit{Hipparcos} catalog, three for every primary source. The primary source is observed three times and each secondary source once, providing a more diverse sidelobe pattern in the \textit{``\texttt{OFF}''} observations. This approach is referred to as ABACAD. In Table \ref{table:obs_table}, we show examples of the two methods\footnote{The naming convention for the ``OFF''  sources from the first strategy shows the name of the primary star with the suffix ``\_OFF".}. Figure \ref{fig:waterfall_HIP4436} shows an example observing set.

With nearly 20\%\ of the observing time on GBT devoted to \BL, observations of the primary target list of $692$ targets in a single receiver band (interspersed with observations of $\sim 2000$ secondary targets) was accomplished in approximately eight months. The future plan of the \BL program is to use additional receivers on the GBT to eventually achieve a full survey coverage from 1 to 12 GHz. Completion of this campaign (1185 stars over the 1$-$12 GHz bandwidth range) is expected to take several years. \\

\subsection{BL Digital Instrumentation}

The \BL digital systems at Green Bank and Parkes are described in detail in \cite{2018PASP..130d4502M} and \cite{PriceBL2017}, respectively. Here we provide a brief summary
of the instrumentation as used in this work. The VEGAS instrument \citep[Versatile Greenbank Astronomical Spectrometer;][]{Prestage2015} is used to digitize and coarsely channelize ($N_{channels} = 512$) one or more dual polarization bands at 3 Gsps (1.5 GHz bandwidth). The digitized voltages are transmitted over an Ethernet network to a cluster of commodity compute servers equipped with multi-TB disk arrays and Graphics Processing Units (GPUs). During observations, channelized voltage
data are written at high speed to local disks, and processed offline using a software spectroscopy suite\footnote{https://github.com/UCBerkeleySETI/gbt\_seti}. This pipeline produces three archival data products: a fine-frequency resolution product dedicated to narrowband spectroscopy (3\,Hz frequency resolution, 18\,s time resolution), a fine-time resolution product dedicated to broadband pulse searches (366\,kHz frequency resolution, 349\,$\mu s$ time resolution), and a mixed product designed for traditional astrophysical investigations (continuum and spectral line; 3\,kHz time resolution, 1\,s time resolution) --- see \cite{LebofskyBL2017} for more details.

The data analysis, described in Sec.~\ref{sec:analysis}, was performed on 796 ABACAD sets, for which the observations had a minimum of three \textit{"\texttt{ON}"} observations from the \textit{A} star. This number is larger than the 692 stars since a subset of the stars were observed on multiple epochs. 
We used the \BL cluster\footnote{A complete description can be found in \cite{2018PASP..130d4502M}} located in Green Bank Observatory for the compute-intensive SETI analysis of this project.  We analyzed 4798 files (180 TB of filterbank data) representing 400\,hr of on-sky time.

\section{Data Analysis}
\label{sec:analysis}

The analysis conducted for this project focused on the detection of narrowband ($\sim$ Hz) signals, potentially drifting in frequency over the duration of an observation.  Spectral drift would be expected due to Doppler shifts from the relative acceleration between transmitter and receiver. This type of signal is of particular interest in traditional SETI projects because it is too narrow to arise naturally from known natural astrophysical sources,
and represents a power-efficient method of transmitting a beacon signal out to great distances. Given the relatively short distances to our targets, we are able to neglect the various interstellar distortions \citep[e.g. scintillation in time and frequency, spectral broadening] {1997ApJ...487..782C,2013ApJ...767...94S}.  We note that our observation planning system requires that observed targets be sufficiently far away from the Sun to allow us to neglect any spectral broadening due to the interplanetary medium.  Thus, to first order, the transmitting frequency of an extraterrestrial continuous radio wave will be affected only by the Doppler acceleration induced by the relative motion between the emitter and the local telescope causing an unknown frequency drift.  The resulting ET waveform $x_{\text{ET}}(t)$ follows \cite{flandrin2001time} and \cite{boashash1992estimating}:
\begin{equation}
x_{\text{ET}}(t)= A e^{i \; 2 \pi \nu(t)\;t} \, ,
\end{equation}
where $A$ is the amplitude, and $\nu(t)$ is referred to as the \emph{instantaneous frequency} of the signal. The waveform is affected by Doppler acceleration by the Earth's (and presumably the hosting system's) orbital and rotational motions (the latter one being the largest contributor). The relative acceleration causes $\nu$ to vary over time in a pseudo-sinusoidal way. Given short observation durations relative to the rotation and orbital periods ($\tau_{\rm{obs}} \approx$ 5~minutes), the change in frequency can be approximated by a linear function, $\nu(t) = \nu_{\rm{ET}}+\dot{\nu}\,t$, with $\nu_{\rm{ET}}$ being the original frequency of the ET transmitter, and $\dot{\nu}$ the shift in frequency (or drift) caused by the Doppler motion. We note that for the narrowband search described here, we replace $\nu_{\rm{ET}}$ with the observation start frequency $\nu_{\rm{0}}$.

The signal detection performance is related to both the energy of the signal $E_{\text{ET}} = A^2$, as well as the frequency drift, as energy gets smeared over the frequency range $\dot{\nu} \times \tau_{\rm{obs}}$  spanned by the signal over the entire observation, where $\dot{\nu}$ is the first-order time-derivative of $\nu(t)$. Uncorrected, the detectability of the signal in the frequency domain decreases proportionally by $\left( \dot{\nu} \times \tau_{\rm{obs}} / \delta\nu\right)^{1/2}$ within a single time-frequency bin, and by a factor of  $\delta t / \tau_{\rm{obs}} $ during the period of the observation, where $\delta t$ is the time resolution.
To maximize the detectability of a received narrowband signal in SETI experiments, a common approach consists of correcting for a set of trial drift rate values, out to a maximum drift rate, and identifying the drift rate value that optimizes detection S/N. This is similar to pulsar and fast radio burst (FRB) searches aimed at maximizing the signal strength for various dispersion measures. As in dedispersion, frequency drift correction can be applied coherently on raw voltage data, or incoherently on detected (total power) spectra. Blind searches over either dispersion or frequency drift generally employ the latter approach.

We have developed a software package, \textsc{\textit{turbo}SETI,}\footnote{\textsc{\textit{turbo}seti}: https://github.com/UCBerkeleySETI/turbo\_seti} that is a Python/Cython implementation of the ``tree deDoppler'' algorithm for incoherent Doppler acceleration searches described in \cite{2013ApJ...767...94S}. This is an extension of the tree search algorithm developed for dispersed pulsar emission searches \citep{1974A&AS...15..367T}. The tree summation algorithm removes redundant operations when summing $n$ spectra over drift paths and reduces the Doppler search algorithm to $\mathcal{O}(n \log n)$ complexity. 

A limit of the incoherent Doppler acceleration search technique is the maximum drift rate before which sensitivity is lost due to energy smearing over adjacent frequency bins during a single time integration. This quantity depends on the size of a single time-frequency pixel. The high-frequency resolution data product produced by the \BL pipeline allows searches of absolute drift rates up to 0.167 Hz s$^{-1}$. The frequency drift induced by Earth's rotation alone is up to 0.16 Hz s$^{-1}$ at 1.4 GHz \citep{1971asee.nasa.....O,2011seti.book.....S}.  This indicates an obvious limitation of the incoherent approach at higher frequencies. 

In \textsc{\textit{turbo}seti}, this limitation is overcome by applying the tree summation to an array that has already been shifted, this allows the search to continue to arbitrarily large drift rates without modifying the frequency resolution of the data \citep[see][for an in-depth discussion]{EnriquezLOFAR2017}. Another solution would be to collapse the data to a lower frequency resolution before applying the tree summation in the algorithm \citep{2013ApJ...767...94S}.

The number of discrete frequency drift rates within a given range that can be searched is a function of the drift rate search resolution.
This in turn depends on $\delta \nu / \tau_{\rm{obs}}$, which corresponds to 0.01 Hz s$^{-1}$ for our high-frequency resolution data products. Thus, a search to a drift rate of $\dot{\nu} = \pm 2$ employs 400 search steps with \textsc{\textit{turbo}seti}.

We perform an analysis on individual 2.9 MHz chunks of spectrum (coarse channels), assuming a uniform gain over the chunk. The RMS noise is evaluated over the fine channels of the zero-drift integrated spectrum. We use the 90th central percentile of the power values to mitigate outliers in the distribution due to the presence of narrowband features and the edges of the poly-phase filterbank response. After each Doppler acceleration correction (or drift rate) the band is summed in time. Any fine-frequency channel that exceeds a minimum signal to noise ratio (S/N) threshold (hereafter a ``hit'') is identified. We define a hit to be the signal with largest S/N at a given frequency channel over all the drift rates searched. The time, frequency, observation meta-data, and a time/frequency subset centered on the hit is recorded to a database for further analysis. 

As a post-processing stage, we remove any hit for which at least one of the ``OFF'' observations has a hit in a range of $\pm 600$\,Hz around the original frequency of the hit. This window corresponds to the maximum frequency change of a signal over the period of the observation given the maximum frequency drift rate searched.

The complete pipeline, including dynamic spectra production, Doppler acceleration correction, and signal detection, has been tested and validated with narrowband anthropogenic extraterrestrial transmissions such as those emitted by the \textit{Voyager 1} spacecraft \citep[][Figure 8.]{2017PASP..129e4501I}.

\section{Results}
\label{sec:results}

We have applied our detection pipeline to approximately 4800 individual, five-minute observations. Using an S/N detection threshold of 20 and a maximum Doppler-drift rate of $\pm \ 2\ $Hz s$^{-1}$ resulted in nearly 29 millions hits.
In post-processing the vast majority of these hits were rejected based on the following criteria.

\begin{enumerate}

\item For the $A$ stars (i.e. ``ON''-source observations), we remove any hit with a drift rate of 0.0 in the topocentric frame. Those signals most likely correspond to ground-based RFI.

\item For the $A$ stars, we only consider hits with an S/N greater than 25. We reserve the S/N range between 20 and 25 for RFI signals, which may potentially be weaker during the ``OFF'' observations, and thus falling below our detection threshold. This attenuation could be expected for a signal that enters through antenna side lobes.

\item Among the remaining hits, we select only those signals present in each of the three $A$ observations. We predict the central frequency of the region where the signal could be located for the immediate following observations by using the drift rate calculated on the first observation. The width of the frequency range used is calculated by using twice the value of the drift rate of the signal. Figure \ref{fig:waterfall_HIP4436} shows an example of such a hit.
\end{enumerate}

\begin{figure}[htb]
\begin{center}
\begin{tabular}{c}
\includegraphics[width=1.0\linewidth]
{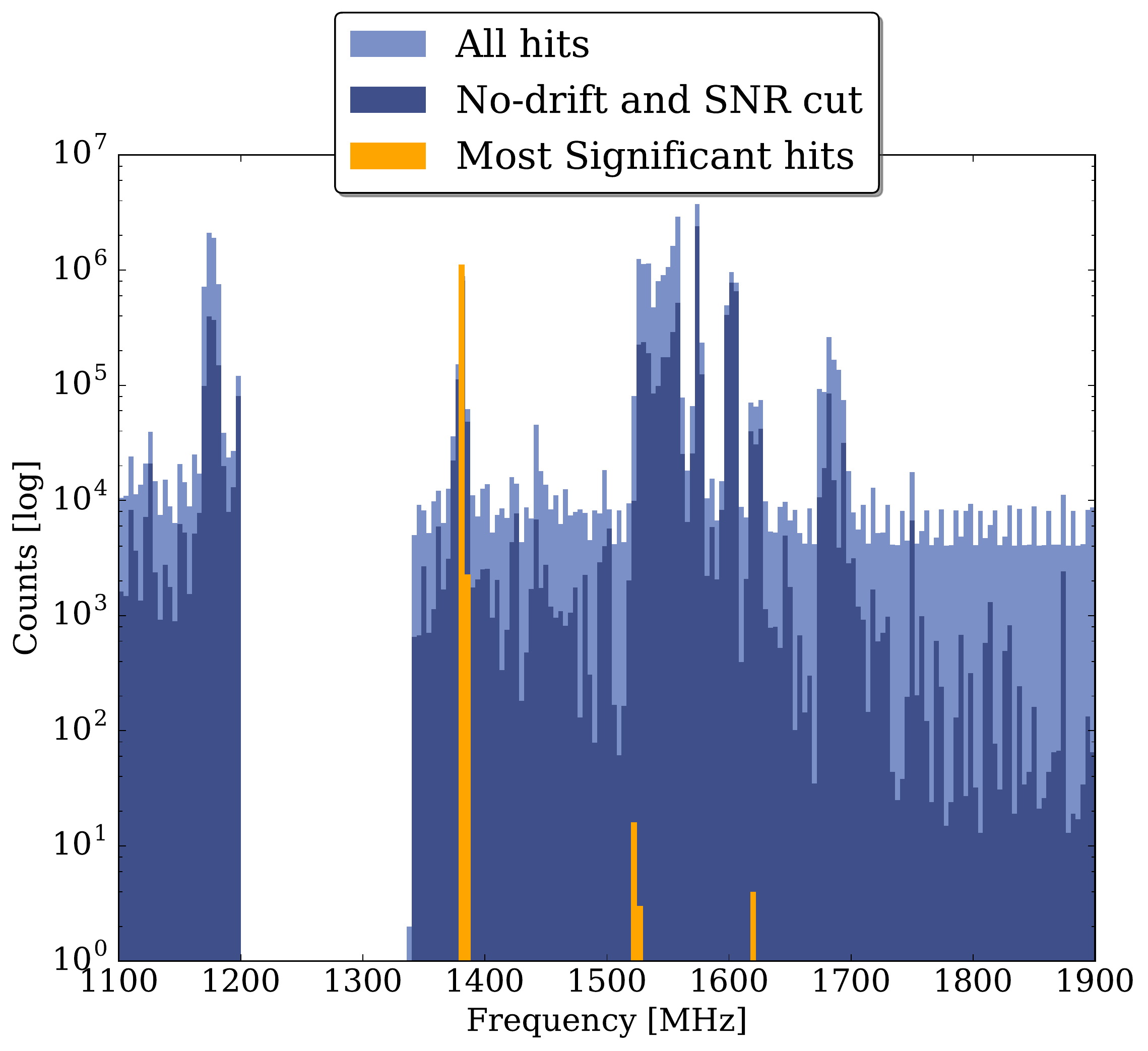}
\end{tabular}
\caption{\footnotesize{Frequency distribution for all the hits produced by the search pipeline (light blue), hits after initial cuts using criteria 1 and 2 from the Results section (dark blue), and the most significant hits that pass all the criteria (orange).}}
\label{fig:rfi_distribution}
\end{center}
\end{figure}

The vast majority of the hits detected in our pipeline can be classified as anthropogenic RFI based on these criteria. Figure \ref{fig:rfi_distribution} shows the frequency distribution of all hits from all observations in this work. There are no hits between 1.2  and 1.35 GHz due to the notch filter. The frequency dependence of the hit distribution is due to the amount of RFI present in those regions of the band. The light blue levels represent the distribution of all 29 million detected hits. The dark blue levels are the hits, which pass criterion 1 and 2 from above. Furthermore, the orange levels are what we determine to be the most significant hits that pass all the criteria.

\subsection{RFI Environment}
\label{sec:rfi}

\begin{figure}[htb]
\begin{center}
\begin{tabular}{c}
\includegraphics[width=1.0\linewidth]
{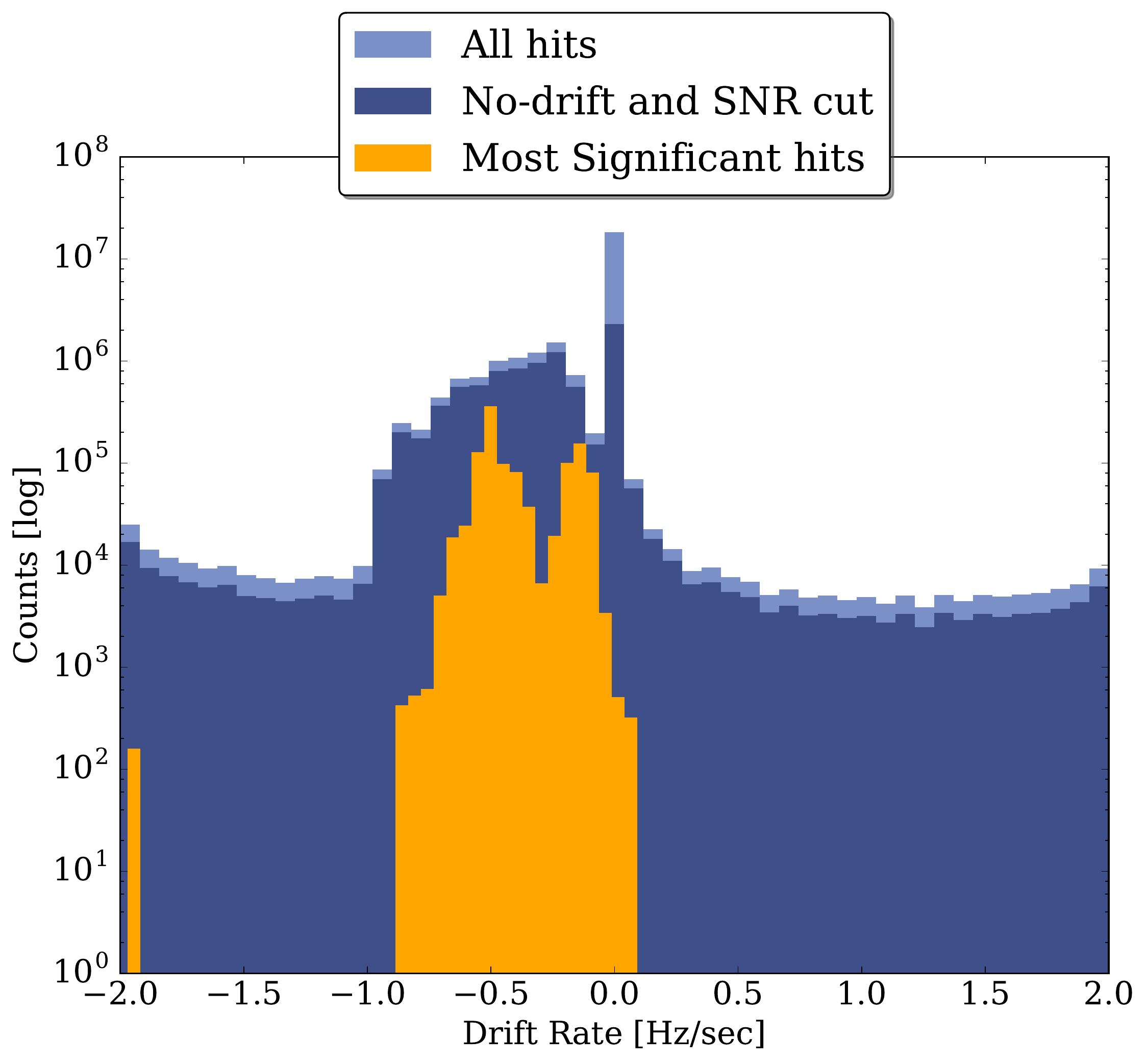}
\end{tabular}
\caption{\footnotesize{ The S/N-maximized drift-rate distribution for the hits. The color scale is described in Figure \ref{fig:rfi_distribution}.}}
\label{fig:drift_distribution}
\end{center}
\end{figure}

The frequency bands allocated for GPS and communication satellites contain the most hits. This is also reflected in Figure \ref{fig:drift_distribution} which shows the distribution of hits as a function of peak drift rate.  A significant increase in the number of hits is observed at negative drift rates, which can be understood to arise from the drift-rate distribution expected from satellites drifting overhead with their acceleration vector pointed toward the center of the Earth.  Stationary RFI signals could appear at any drift rate (e.g. sweeping transmissions or instrumental artifacts), but most stationary terrestrial narrowband interferers, without intrinsic frequency modulation, would show no measurable drift.  From Figure \ref{fig:drift_distribution}, we can see that these zero-drift interferers are the most common type detected by our pipeline.

\subsection{Most Significant Events}

Our significance criteria filter results in 11 ``events'' which required further analysis to classify. We define ``events'' as one or more hits during observations of a single star system in a single epoch. These observations and detections are listed in Table \ref{table:hits_table}. 
We have listed the source, the observation date and starting time, the frequency of the detected signal based on the beginning of the observation, the S/N-maximized drift-rate, and the S/N for each of these events.  Upon further analysis, we can classify each of these events as likely associated with a terrestrial source.
 
\begin{table}
\centering
\caption{Most Significant Events that Pass our Detection Criteria.}
\begin{tabular}{lccccc}
\hline
\T Source & Decimal MJD & Frequency  & Drift Rate & S/N \\
\B   &   & (MHz)  &  (Hz s$^{-1}$) &   \\

\hline
\hline
HIP 17147  & 57523.802997685183 & 1379.27751  &  $-0.266$  &  25.4\\
HIP 4436   & 57803.934409722220 & 1380.87763  &  $-0.507$  &  463.3\\
HIP 20901  & 57606.579375000001 & 1380.97122  &  $-0.478$  &  84.6\\
HIP 39826  & 57456.030891203707 & 1380.92937  &  $-0.542$  &  420.3\\
HIP 99427  & 57752.960949074077 & 1380.92570  &  $-0.086$  &  50.2\\
HIP 66704  & 57650.631631944445 & 1380.91201  &  $-0.134$  & 3376.9\\
HIP 82860  & 57664.923159722224 & 1381.20557  &  $-0.335$  &  435.4\\
HIP 74981  & 57523.259328703702 & 1384.20759  &  $-0.246$  &  237.7\\
HIP 65352  & 57459.396956018521 & 1522.18102  &  $+0.010$  &  113.6\\
HIP 45493  & 57636.782812500001 & 1528.46054  &  $-0.010$  &  32.1\\
HIP 7981   & 57680.179629629631 & 1621.24028  &  $+0.660$  &  38.7\\ \hline
\label{table:hits_table}
\end{tabular}
\textbf{Note.} For each event, the source at boresight, observation date, frequency, S/N-maximized drift-rate, and S/N are listed.

\end{table}

Eight of these events have multiple hits (in some cases, up to hundreds of thousands over the three observations); for brevity, we only report the highest S/N hit in Table~\ref{table:hits_table}. Complete information on all hits can be found on the survey website.

An example of one of these events is shown in Figure~\ref{fig:waterfall_HIP4436}, illustrating the detection of a strong hit at around 1380.87\,MHz. The signal can be seen drifting toward lower frequencies in the following two ``ON'' observations. This is, in essence, a type of signal we would expect from an extraterrestrial transmitter affected by the acceleration of both the host planet and the Earth. This type of signal is correctly reported as a possible detection by our pipeline. However, we discount the signal as extraterrestrial for the following reasons.

These eight events show similar morphology, in particular, many hits with a wide range of drift rates. 
Moreover, all the hits from these events have similar frequencies around 1380 MHz, which is often used for long-range air traffic control (ATC) radar and GPS, among other uses\footnote{https://www.ntia.doc.gov}.
These characteristics lead us to believe that the signals are unlikely to be originating outside the solar system. 

\begin{figure}[h!]
\begin{center}
\begin{tabular}{c}
\includegraphics[width=1.0\linewidth]
{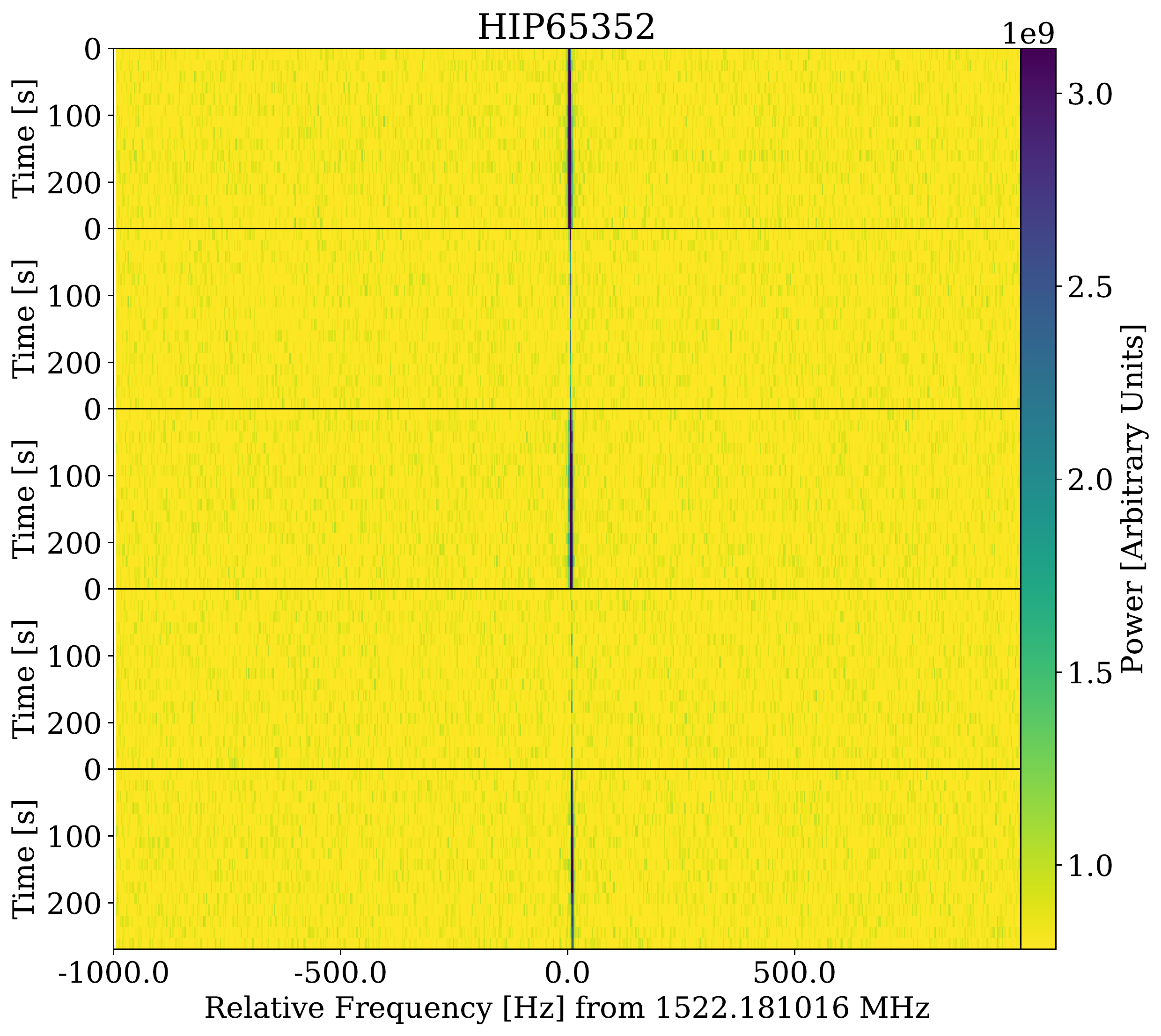}
\end{tabular}
\caption{\footnotesize{Series of five minute "ON-OFF" observations of HIP 65352 as described in Figure \ref{fig:waterfall_HIP4436}. This was reported as a significant event because the weaker signal in the ``OFF'' observation was not detected by the pipeline. 
}}
\label{fig:waterfall_HIP65352}
\end{center}
\end{figure}

Another two of the events were found during observations of HIP\,65352 and HIP\,45493. They contain hits at the minimum drift rate of $\pm 0.1$\,Hz s$^{-1}$ and are both at frequencies of $\sim1520$ MHz. Figure \ref{fig:waterfall_HIP65352} shows the presence of the signal during the ``OFF'' observations, although much weaker. These ``OFF'' signals are slightly below our initial detection threshold, and thus are not reported.  The presence of the signal in the ``OFF'' observations indicates this emission is coming from a nearby stationary source.

\begin{figure}[htb]
\begin{center}
\begin{tabular}{c}
\includegraphics[width=1.0\linewidth]
{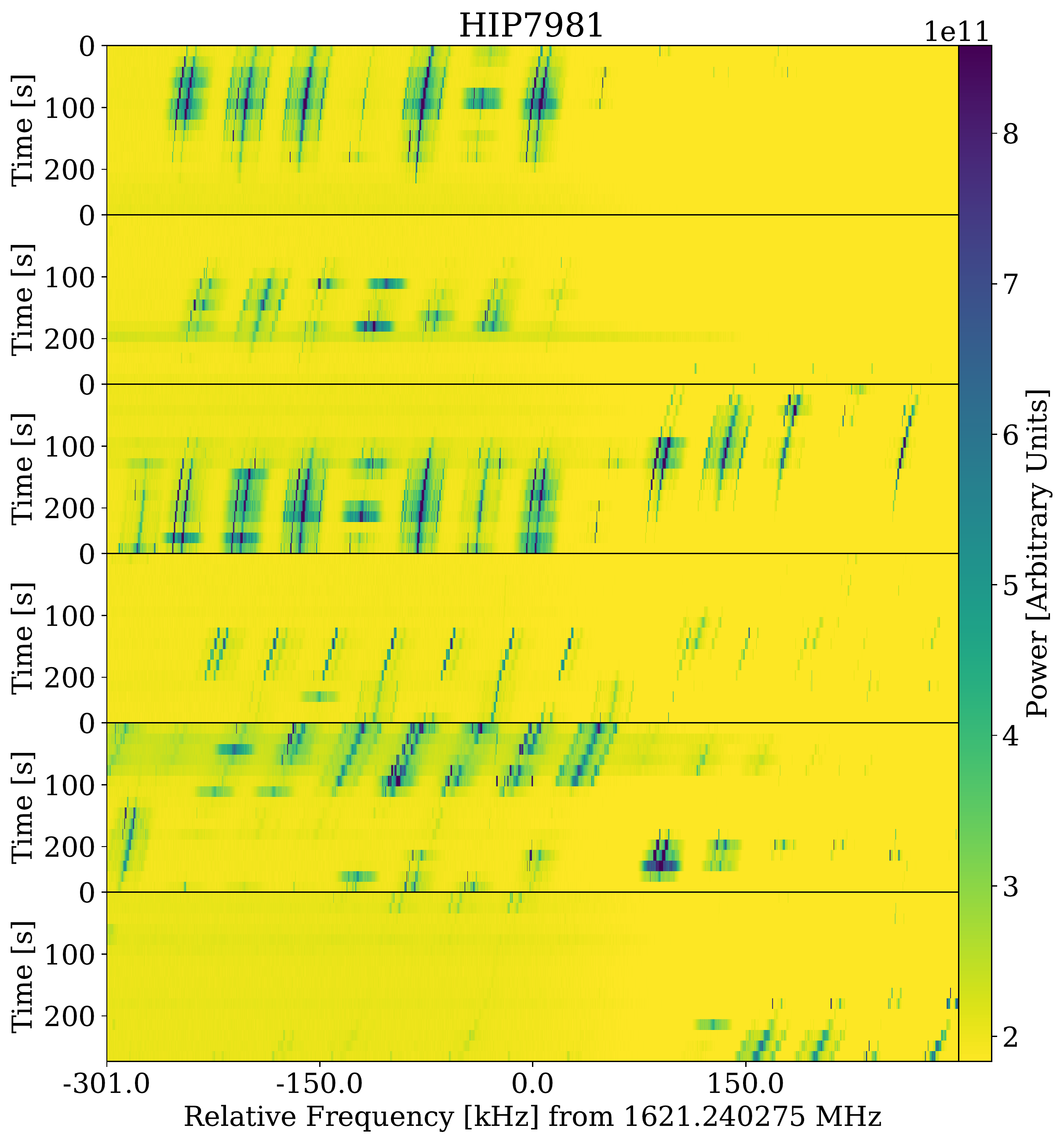}
\end{tabular}
\caption{\footnotesize{Series of five minute "ON-OFF" observations of HIP 7981 as described on Figure \ref{fig:waterfall_HIP4436}. The complex structure appears in both ``ON'' and ``OFF'' observations.}}
\label{fig:waterfall_HIP7981}
\end{center}
\end{figure}

The last event, detected while observing HIP\,7981, is unique.
It has a moderate drift rate ($+0.66$), S/N (38.7) and is at a different frequency compared to other false-positive events. However, upon visual inspection (see Figure ~\ref{fig:waterfall_HIP7981}) there is a complex structure across the band, a higher drift-rate search would result in a higher S/N detection, and a similar morphology of the signal can be seen in all of the ``OFF'' observations. We are unsure what this complex signal source is, but we consider it anthropogenic due to its presence in independent pointings.

We conclude that the 11 significant events reported by our detection pipeline are the types of signals we expect to detect based on our observation strategy, observing band, and detection pipeline. However, we can state with high certainty that these events are false-positives that were initially detected as significant due to the complex and varied nature of anthropogenic RFI.

We are continually improving our detection pipeline to be able to set lower detection thresholds without significantly increasing the number of false-positive events, or computational load. Future versions of our detection pipeline are being designed to successfully filter events such as these.

\section{Discussion}
\label{sec:discussion}

For a signal to be attributable to extraterrestrial technology, it must be clear that the signal was neither generated by astrophysical processes nor by a human-made transmitter. For this reason, SETI searches often implement spectrometers with very narrow channel bandwidths ($\sim$ Hz resolution), which provide fine spectral detail. Furthermore, signal detectability reduces according to the frequency resolution and signal bandwidth mismatch. The data analysis presented in this paper focuses on narrowband signals. We aim to address other signal types -- in particular, pulsed broadband signals -- in future detection pipelines employing a wider variety of signal detection methodologies \citep[e.g.][]{2015aska.confE.116S}. For example, the signal found while observing HIP\,7981 could potentially be identified by a machine learning (ML) approach as local RFI.

\subsection{Sensitivity and Transmitter Power}
\label{sec:sens}

The sensitivity of a radio-frequency SETI experiment is determined primarily by the system noise and effective collecting area of the telescope, which can be encapsulated in the system equivalent flux density (SEFD):
\begin{equation}
\mathrm{SEFD} = \frac{2k_{\rm{B}}T_{\rm{sys}}}{A_{\rm eff}} \, ,
\end{equation}
where $k_{\rm{B}}$ is the Boltzmann constant and $T_{\rm{sys}}$ is the system temperature due to various sources of noise. The effective collecting area,  $A_{\rm eff}=\eta A$, where $A$ is the physical collecting area of the telescope and $\eta$ is an efficiency factor between 0 and 1. The SEFD is reported in Jy (1~Jy = 10$^{-26}$ W m$^{-2}$ Hz$^{-1}$). The fraction $A_{\rm eff}/2k_{\rm{B}}$ is also known as the telescope gain factor $G$ (units K Jy$^{-1}$) which can be determined by observing calibrator sources.
For the GBT at L band, the SEFD is approximately 10~Jy \footnote{https://science.nrao.edu/facilities/gbt/proposing/GBTpg.pdf/view}.

For observations of astrophysical sources, the astrophysical signal is generally wider than the frequency resolution of the measurement. For those cases, the minimum detectable flux density $S_{\rm{min}}$ is given by
\begin{equation}
S_{\rm{min, wide}} = \rm{S/R}_{\rm{min}} \; \frac{\rm{SEFD}}{\sqrt{n_{pol}\Delta\nu ~\tau_{\rm{obs}}}} \, ,
\label{eq:sensitivity_astro}
\end{equation}
where $\rm{S/N}_{\rm{min}}$ is a signal-to-noise threshold value,  $\tau_{\rm{obs}}$ is the observing time,  $\Delta\nu$ is the bandwidth, and $n_{\rm{pol}}$ is the number of polarizations.
However, in the case of extremely narrowband signal detection (i.e. the transmitter signal bandwidth is narrower or equal to the observing spectral resolution) the minimum detectable flux density $S_{\rm{min}}$ is then given by
%

\begin{equation}
S_{\rm{min, narrow}} = \rm{S/N}_{\rm{min}} \; \frac{\rm{SEFD}}{\delta\nu_{\rm{t}}}\sqrt{\frac{\delta\nu}{n_{pol}\tau_{\rm{obs}}}} \, ,
\label{eq:sensitivity}
\end{equation}
where $\delta\nu$ is the observing channel bandwidth and $\delta\nu_{\rm{t}}$ is the bandwidth of the transmitting signal.
Assuming an SEFD of 10\,Jy\footnote{The GBT L-band receiver is sufficiently stable that we can use this estimate as a consistent conservative value \citep{2011A&A...536A..81B}.} across the band, the minimum detectable flux density for a five-minute L-band observation with the GBT, at 3 Hz resolution for an S/N of at least 25 is 17 Jy. \\ 
\\
Using this sensitivity, we can set a minimum luminosity (transmitter power) detection threshold based on the distance to each system observed. The intrinsic
luminosity $L$ of a source is
\begin{equation}
L=4\pi d_{\star}^{~2}S\, ,
\end{equation}

where $d_{\star}$ is the distance to the source, and $S \gtrsim S_{\rm{min}} $.  For a distance of 10, 100, and 1000 lt-yr the minimum detectable luminosity is 28 GW, 2.8 TW, and 280 TW
respectively. These are very large power requirements, but assuming a high-gain antenna with a transmitter pointed at Earth, the power requirement is significantly reduced. We can
associate the power of the transmitter $P_{\rm{tx}}$ with the detected flux density by setting the luminosity to be equal to the Equivalent Isotropic
Radiated Power (EIRP) of an antenna:
\begin{equation}
\mathrm{EIRP} = G_{\rm{ant}} P_{\rm{tx}}\, ,
\end{equation}
where $G_{\rm{ant}}$ is the antenna gain relative to an idealized isotropic antenna. In this context, the luminosity and the EIRP are equivalent, resulting in
\begin{equation}
S = \frac{G_{\rm{ant}} P_{\rm{tx}}}{4 \pi d_{\star}^2}\, ,
\end{equation}
The gain of a parabolic radio antenna with diameter, $D$, is given by
\begin{equation}
G_{\rm{parabolic}} = \frac{4 \pi A_{\rm{eff}}}{\lambda^2} = \epsilon \left(\frac{\pi D}{\lambda}\right)^{2} \, ,
\end{equation}
where $\epsilon$ is the measured telescope efficiency factor, and $\lambda$ is the
observing wavelength.  

Using the Arecibo dish as a fiducial high-gain antenna, the gain of which is approximately $4.3 \times 10^{7}$ at L band, results in a minimum power requirement of 650 W, 65 kW, 6.5 MW (at distances of 10, 100, 1000 lt-yr) under the ideal situation in which both the transmitting and receiving telescopes are aligned. All stars in the observed sample are within 50 parsecs ($\sim163$ lt-yr). In the ideal case of a planetary radar system similar to Arecibo transmitting continuously at Earth, our survey is sufficiently sensitive to detect such a signal from any of the observed star systems in our survey.

\begin{turnpage}
\begin{table*}
\centering
\caption{Selected Searches: First Part of Table Showing the Parameters Used for Different Searches.}

\begin{tabular}{l|c|cc|cccc|c|ccccc}
\hline 
\B\T & \textbf{This work ~\footnote{Unless specified otherwise, most values fro the GBT are taken from \url{https://science.nrao.edu/facilities/gbt/proposing/GBTpg.pdf/view}}} & \multicolumn{2}{c|}{\textbf{Gray 2017}} & \multicolumn{4}{c|}{\textbf{Harp 2016}} & \textbf{Siemion 2013} & \multicolumn{5}{c}{\textbf{Project Phoenix ~\footnote{Most values taken form \cite{2016AJ....152..181H}, references therein, as well as a private communication with Gerry Harp and Jill Tarter, unless otherwise specified.}}}\tabularnewline
\hline 
\hline
\B\T \textbf{TELESCOPE PARAMETERS ~\footnote{Most information in this table comes from \citep{2013ApJ...767...94S,2016AJ....152..181H,2017AJ....153..110G}}}  &  & \multicolumn{2}{c|}{} & \multicolumn{4}{c|}{} &  & \multicolumn{5}{c}{}\tabularnewline
\G Telescope(s) & GBT & \multicolumn{2}{c|}{VLA} & \multicolumn{4}{c|}{ATA} & GBT & \multicolumn{2}{c}{AO} & \multicolumn{2}{c}{Parkes} & NRAO 140'\tabularnewline
Antenna Diameter ($D$) {[}m{]} & 100~\footnote{The dimensions of GBT are 100m x110m. However, we used 100m here.%
} & \multicolumn{2}{c|}{25} & \multicolumn{4}{c|}{6.1} & 100~\ft{$^{\rm{d}}$} & 305 & 225 & \multicolumn{2}{c}{64} & 43\tabularnewline
\G Number of Antennas per Telescope & 1 & \multicolumn{2}{c|}{27} & \multicolumn{4}{c|}{27} & 1 & \multicolumn{5}{c}{1}\tabularnewline
Beam Width [arcmin]~\footnote{Calculated using the central frequency.} & 8 & \multicolumn{2}{c|}{57~\footnote{We quote here the image size. The FWHP beam with is 32' at 1.4 GHz \citep{2017AJ....153..110G}.}} & \multicolumn{4}{c|}{$3\times6~\footnote{We note this is the value calculated for 1.4 GHz.}$} & 8 & 3 & 2 & 13 & 8 & 14\tabularnewline
\G Aperture Efficiency ($\eta$) & 0.72 & \multicolumn{2}{c|}{0.45~\footnote{From \cite{2009IEEEP..97.1448P}}} & \multicolumn{4}{c|}{0.58~ \footnote{Expected value at 1.5GHz; taken from Welsh \& DeBoer (2004): \url{http://www.seti.org/sites/default/files/ATA-memo-series/memo66.pdf}}} & 0.72 & 0.7 & 0.7 & 0.7 & 0.7 & 0.7 \tabularnewline
System Temperature ($T_{\mathrm{sys}}$) {[}K{]} & 20 & \multicolumn{2}{c|}{35} & \multicolumn{4}{c|}{108~\footnote{Average value calculated from the values published in \cite{2016AJ....152..181H}.}} & 20 & 40 & 40 & 35 & 35 & 35\tabularnewline
\hline 
\B\T \textbf{SEARCH PARAMETERS ~\ft{$^{\rm{c}}$}} &  & \multicolumn{2}{c|}{} & \multicolumn{4}{c|}{} &  & \multicolumn{5}{c}{}\tabularnewline
\G Number of stars & 692 & \multicolumn{2}{c|}{$10^{12}$} & 65 & 1,959 & 2,822 & 7,459 & 86 & 290 & 371 & 206 & 105 & 195\tabularnewline
Distance to Stars [pc(Ly)]%
&\specialcell{ 50\\ (163)} & \multicolumn{2}{c|}{ \specialcell{$7.8\times10^{5}$ \\($2.5\times10^{6}$)} } & \specialcell{1,400~\footnote{Distance to Kepler30, the maximum distance found for this group.} \\(4,566)} 
 & \specialcell{1,000~\footnote{For these surveys, we have adopted 1kpc as a characteristic distance.} \\(3,200)} & \specialcell{300~\footnote{From \cite{2003ApJS..145..181T}} \\ (978) }  & \specialcell{500~\footnote{From \cite{2003ApJS..149..423T}} \\ (1630)} & \specialcell{1,000~\ft{$^{\rm{l}}$}\\ (3,200)} & \multicolumn{5}{c}{\specialcell{215~\footnote{From \cite{2000AcAau..46..649S}.} \\ (700)}}\tabularnewline
\G Stellar Spectral Types & BAFGKM & \multicolumn{2}{c|}{All} & FGK~\footnote{Distribution taken from the Kepler mission star distribution.} & FGK~\ft{$^{\rm{p}}$} 
& FGKM~\ft{$^{\rm{m}}$} & BAFGKM & 
FGK~\ft{$^{\rm{p}}$}
 & \multicolumn{5}{c}{FGK}\tabularnewline
S/N Threshold & 25 & \multicolumn{2}{c|}{7} & \multicolumn{4}{c|}{6.5~ \footnote{Originally, an S/N of 9 and integration time of 192 s was used. These were later changed to S/N of 6.5 and integration time of 93 s. It is not clear when this change happened, but the sensitivity value is about the same for both configurations.  
}} & 25 & \multicolumn{5}{c}{\textbf{...}}\tabularnewline
\G Spectral Resolution ($\delta\nu$) {[}Hz{]} & 3 & 122 & 15.3 & \multicolumn{4}{c|}{0.7} & 1 & \multicolumn{5}{c}{1.0 ~\footnote{From \cite{2002ASPC..278..525B}.}}\tabularnewline
Frequency Coverage {[}GHz{]} & 1.1--1.9 & \multicolumn{2}{c|} {1.4$\pm$0.001} 
& \multicolumn{4}{c|}{1--9} & 1.1--1.9 & 1.2--1.75 & 1.75--3.0 & 1.2--1.75 & 1.75--3.0 & 1.2--3.0 \tabularnewline
\G Total Bandwidth ($\Delta\nu_{\rm{tot}}$) {[}MHz{]} & 660 & 1 & 0.125 & 8000 & 2040 & 337 & 268 & 670 & 370~\ft{$^{\rm{q}}$} 
& 1,250 & 550 & 1,250 & 1,800\tabularnewline
Instantaneous Bandwidth ($\Delta\nu_{\rm{obs}}$) {[}MHz{]} & 800~\footnote{This is the total instantaneous band recorded. In post-processing, we removed 140 MHz of bandwidth, which is suppressed by a notch filter.} & 1 & 0.125 & \multicolumn{4}{c|}{70} & 670 & \multicolumn{5}{c}{20}\tabularnewline
\G Central Frequency ($\nu_{\rm{mid}}$) {[}GHz{]} & 1.5 & 1.4 & 8.4 & \multicolumn{4}{c|}{5.0} & 1.5 & 1.5 & 2.375 & 1.5 & 2.375 & 2.1\tabularnewline
Time Resolution ($\delta t$) {[}s{]} & 18 & 5 & 5 & \multicolumn{4}{c|}{1.5} & 1 & \multicolumn{5}{c}{\textbf{...}}
\tabularnewline
\G Total Integration Time ($\tau_{\rm{obs}}$) {[}s{]} & 300 & 1,200 & 300 & \multicolumn{4}{c|}{93~\ft{$^{\rm{q}}$}}
& 300 & 276 & 195 & 276 & 138 & 552\tabularnewline
\hline 
\B\T \textbf{CALCULATED PARAMETERS}  &  &  &  &  &  &  &  &  &  &  &  &  & \tabularnewline
\G SEFD {[}Jy{]} & 10 & \multicolumn{2}{c|}{18} & \multicolumn{4}{c|}{664} & 10 & 
 2.2 &  2.2 & 42.9 &  42.9 & 99.65
\tabularnewline
Sensitivity~
\footnote{We assume the original signal would be 1 Hz wide. We ignore the various Doppler acceleration correction technique used.} {[}Jy{]} 
& 17 & 28 & 20 & \multicolumn{4}{c|}{378} & 10 & 16~\ft{$^{\rm{u}}$} & 16~\ft{$^{\rm{u}}$} & 100~\ft{$^{\rm{u}}$} & 100~\ft{$^{\rm{u}}$} & 100~
\footnote{Values taken from \cite{2016AJ....152..181H}. These values were used in the figure-of-merit calculations.}
\tabularnewline
\G EIRP {[}W{]} & $5.2\times10^{12}$ & $2.0\times10^{21}$ & $1.4\times10^{21}$ & 
$8.8\times10^{16}$ & $5.5\times10^{16}$ & $3.8\times10^{14}$ & $1.1\times10^{15}$
& $1.4\times10^{15}$ & 
$8.8\times10^{13}$ & $8.8\times10^{13}$ & $5.5\times10^{14}$ & $5.5\times10^{14}$ & $5.5\times10^{14}$
\tabularnewline
Sky Coverage {[}deg$^{2}${]} & 10.6 & \multicolumn{2}{c|}{22.7} & 
\multicolumn{4}{c|}{193}
& 1.3 & 
\multicolumn{5}{c}{18}
\tabularnewline
\G CWTFG & 0.85 & \multicolumn{2}{c|}{136.3} & \multicolumn{4}{c|}{2970} &1878 & \multicolumn{5}{c}{49}   \tabularnewline
\hline
\end{tabular}

\label{table:selected_searches}

\textbf{Notes.} This part shows some of the most modern SETI searches, and well as the Phoenix project. This is the arXiv version of the table. It comprises Tables 3 and 4 shown in the ApJ version.

\end{table*}

\end{turnpage}

\begin{table*}
\centering
\caption{Selected Searches : Second part of Table Showing the Parameters Used for Different Searches.}
\begin{tabular}{l|c|cc|c|cc}
\hline 
\B\T & \textbf{Horowitz 1993} & \multicolumn{2}{c|}{\textbf{Valdes 1986}} & \textbf{Tarter 1980} & \multicolumn{2}{c}{\textbf{Verschuur 1973}}\tabularnewline
\hline 
\hline 
\rule{0pt}{3.6ex} \rule[-2.2ex]{0pt}{0pt} \textbf{TELESCOPE PARAMETERS~ \footnote{Most information in this table comes from \citep{1973Icar...19..329V,1980Icar...42..136T,1986Icar...65..152V,1993ApJ...415..218H}. When different specs were used during an experiment, We have taken the most optimistic values for each.}} &  & \multicolumn{2}{c|}{} &  & \multicolumn{2}{c}{}\tabularnewline
\G Telescope(s) & Harvard-Smithsonian 26m & \multicolumn{2}{c|}{HCRO 26m} & NRAO 91m & NRAO 300' & NRAO 140'\tabularnewline
Antenna Diameter ($D$) {[}m{]} & 26 & \multicolumn{2}{c|}{26} & 91 & 91 & 43\tabularnewline
\G Number of Antennas per Telescope & 1 & \multicolumn{2}{c|}{1} & 1 & 1 & 1\tabularnewline
Beam Width [arcmin]~\footnote{Calculated using the central frequency.} & 30 & \multicolumn{2}{c|}{32} & 8 & 10 & 21 \tabularnewline
\G Aperture Efficiency ($\eta$) & 0.5 \footnote{We were unable to find a value in the literature. We assume a similar value to the antenna of the same dimensions from \cite{1986Icar...65..152V}.} & \multicolumn{2}{c|}{0.5} & 0.6 & 0.6 & 0.6~
\footnote{This value was taken from the NRAO 300 feet for our calculations, we were unable to find a value in the literature for the 140 feet.
}\tabularnewline
System Temperature ($T_{\mathrm{sys}}$) {[}K{]} & 85 & \multicolumn{2}{c|}{100} & 70 & 110 & 48\tabularnewline
\hline 
\B\T \textbf{SEARCH PARAMETERS ~\ft{$^{\rm{a}}$}} &  & \multicolumn{2}{c|}{} &  &  & \tabularnewline
\G Number of stars & $10^{7}$ ~\ft{$^{~\rm{e}}$}
& 53 & 12 & 201 & 3 & 8\tabularnewline
Distance to Stars [pc(Ly)]%
& 700~\footnote{\cite{1993ApJ...415..218H} suggested values for the number of stars given a distance, based on the power of an isotropic beacon.} (2283) & \multicolumn{2}{c|}{ 6.1 (20)} & 25(82) & \multicolumn{2}{c}{5(16)}\tabularnewline
\G Stellar Spectral Types & All & \multicolumn{2}{c|}{BFGKM~%
\footnote{The variety of targets in this project was very heterogeneous. It included stars, galaxies, pulsars, and even planets. Only the stellar sources were used when compared to this work.%
}} & FGK & \multicolumn{2}{c}{GKM}\tabularnewline
S/N Threshold & 30 & \multicolumn{2}{c|}{3.0} & 12 & \multicolumn{2}{c}{3~
\footnote{It was only specified that the data were \textquotedbl{}inspected\textquotedbl{}.
Thus we assume a 3$\sigma$ threshold.%
}}\tabularnewline
\G Spectral Resolution ($\delta\nu$) {[}Hz{]} & 0.05 & 4,883 & 76 & 5.5 & 490 & 7,200\tabularnewline
Frequency Coverage {[}GHz{]} & $1.4200\pm0.0002$ & \multicolumn{2}{c|}{$1.5167\pm0.0007$} & $1.6664\pm0.0007$ & \multicolumn{2}{c}{$1.426\pm0.010$}\tabularnewline
\G Total Bandwidth ($\Delta\nu_{\rm{tot}}$) {[}MHz{]} & 1.2 & 1.25 & 0.078 & 1.4 & 0.6 & 20\tabularnewline
Instantaneous Bandwidth ($\Delta\nu_{\rm{obs}}$) {[}MHz{]} & 0.4 & 1.25 & 0.078 & 0.36 & 0.6 & 2.5\tabularnewline
\G Central Frequency ($\nu_{\rm{mid}}$) {[}GHz{]} & 1.42 & \multicolumn{2}{c|}{1.5167} & 1.6664 & \multicolumn{2}{c}{1.426}\tabularnewline
Time Resolution ($\delta t$) {[}s{]} & 20 & \multicolumn{2}{c|}{300} & 0.2 & \multicolumn{2}{c}{10}\tabularnewline
\G Total Integration Time ($\tau_{\rm{obs}}$) {[}s{]} & 20 & \multicolumn{2}{c|}{3000} & 45 & 240 & 300\tabularnewline
\hline 
\B\T \textbf{CALCULATED PARAMETERS} &  &  &  &  &  & \tabularnewline
\G SEFD {[}Jy{]} & 884 & \multicolumn{2}{c|}{1040} & 51 & 62 & 124\tabularnewline
Sensitivity~ 
\footnote{We assume the original signal would be 1 Hz wide. We ignore the various Doppler acceleration correction techniques used.} {[}Jy{]} & 18,755& 3980  & 497 & 150 & 187 & 1284 \tabularnewline
\G EIRP {[}W{]} & $1.1\times10^{18}$ & $1.8\times10^{13}$ & $2.2\times10^{12}$ & $1.1\times10^{13}$ & $5.6\times10^{11}$ & $3.8\times10^{12}$\tabularnewline
Sky Coverage {[}deg$^{2}${]} & 28,052 & \multicolumn{2}{c|}{14.7} & 3.0 & \multicolumn{2}{c}{1.6}\tabularnewline
\G CWTFG & 6506 & \multicolumn{2}{c|} {20,208} & 3233 & \multicolumn{2}{c} {1,693} \tabularnewline
\hline 
\end{tabular}

\label{table:selected_searches2}

\textbf{Note.} This part shows some of the early searches spanning the first couple of decades of SETI.
\end{table*}

\subsection{Figures of Merit}

\begin{figure*}[htb]
\begin{center}
\begin{tabular}{c}
\includegraphics[width=0.8\linewidth]
{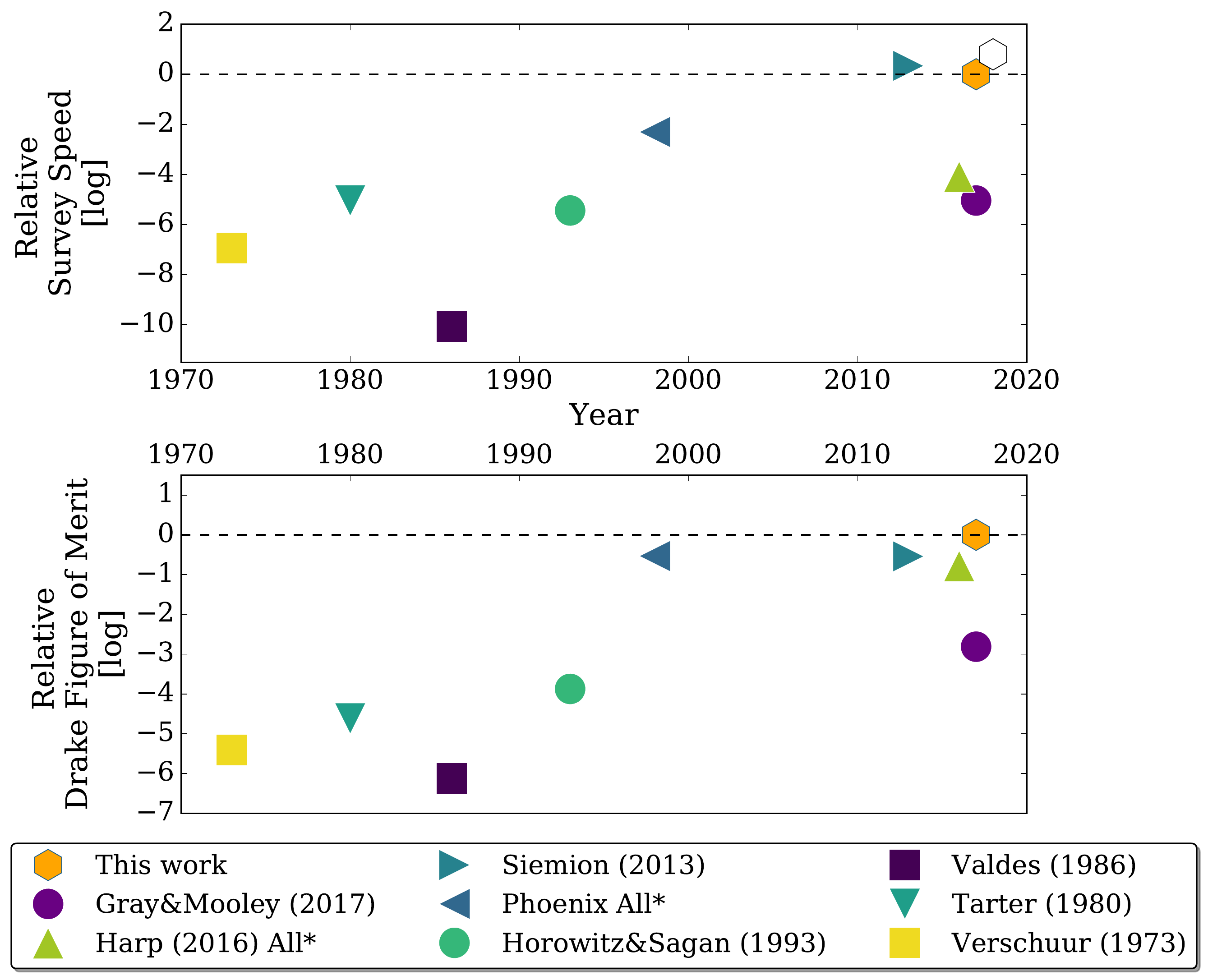}
\end{tabular}
\caption{Comparison of this work with several previous SETI campaigns. The top figure compares surveys based on relative survey speeds. The white hexagon takes into account the current instantaneous bandwidth available to the \BL backend ($\Delta\nu \approx$ 5 GHz), which is underutilized in L band observations.  The bottom figure uses the relative DFM values for the comparison. Both figures only show the summed values for surveys with multiple components. The values used to make this figure can be found in Tables \ref{table:selected_searches} and \ref{table:selected_searches2}.}
\label{fig:other_merit}
\end{center}
\end{figure*}

The unknown nature and characteristics of a putative ET signal creates a large parameter space that one needs to search. This, in general, makes the comparison of SETI surveys challenging.
Previous studies have calculated figures-of-merit for comparison. These figures-of-merit vary wildly and often depend acutely on what the authors think are the most important parameters.
In this section, we describe several different figures-of-merit in order to show multiple perspectives, as well as to provide context to our work with respect to previous studies\footnote{Note that all the values used for the Figures \ref{fig:other_merit} and \ref{fig:galactic_rarity} are shown in Tables \ref{table:selected_searches} and \ref{table:selected_searches2}.}.

We have endeavored to include all significant radio SETI surveys in this section, but some surveys have not been sufficiently reported in the astrophysical literature, or are sufficiently different in the sampling of the parameter space, so that a comparison is difficult (e.g. SERENDIP, SETI@home, \cite{1961PhT....14...40D}).

\subsubsection{Survey Speed}

Survey speed is a standard figure-of-merit used in radio astronomy surveys to describe the efficiency of surveys in relation to the telescope and instrumentation used.
Assuming a survey conducted for a given sensitivity $S_{\rm{min}}$ and threshold $S/N_{\rm{min}}$, the speed at which such a search can be  completed depends on the SEFD and instantaneous bandwidth covered ($\Delta\nu_{\rm{obs}}$). 
Thus, a Survey Speed Figure-of-Merit (SSFM) can be defined as
\begin{equation}
 SSFM \propto \frac{\Delta\nu_{\rm{obs}}}{\rm{SEFD}^2 ~\delta\nu} \, ,
\label{eq:SSFM}
\end{equation}
The upper panel of Figure \ref{fig:other_merit} shows the relative SSFM for several SETI efforts. The values were calculated by normalizing them to the \BL SSFM, thus for slower surveys the relative SSFM $<1$.

Relative speed is important; it shows in this case that our search is millions of times faster than some of the very early searches, making our search previously infeasible \footnote{This was already noted during the Phoenix project, \cite{2000ASPC..213..451C} noted that it would take thousands of years to observe millions of stars. At speeds of soon available facilities, this could be done in less than a decade}. However, this figure-of-merit lacks the ability to compare the full extent of individual targeted programs, neglecting information about the number and types of targets observed.

\subsubsection{The Drake Figure of Merit}

One of the most well-known figures-of-merit in the SETI literature is the Drake Figure-of-Merit (DFM) \citep{1983sswg.book.....D}. It is commonly defined as
\begin{equation}
DFM = \frac{\Delta\nu_{\rm{tot}}~\Omega}{~~\rm{S_{\rm{min}}}^{3/2}}\, ,
\label{eq:DFM}
\end{equation}
where $\Delta\nu_{tot}$ is the total bandwidth and $\Omega$ is the total sky coverage.
The lower panel of Figure \ref{fig:other_merit} shows the relative DFM for the same set of SETI projects.  Numerical values were calculated normalized to the \BL DFM.

The DFM is able to compare searches over their total parameter space searched in terms of sky coverage and frequency coverage. However, it gives equal weight to any part of the sky, assuming an isotropic distribution of ET transmitters. One could argue that an observation pointed toward a known star, galaxy, or the center of the Milky Way, would have more value than ``empty'' regions of the sky.

\subsubsection{Other Figures of Merit}

One example of a figure-of-merit developed in \cite{2016AJ....152..181H} uses $N_{\rm{stars}} \times \Delta\nu_{\rm{tot}}$
, where $N_{\rm{stars}}$ is the total number of stars observed and $\Delta\nu_{\rm{tot}}$ is the total bandwidth covered. Unfortunately, this does not take into account the sensitivity of an observation making it difficult to compare searches using telescopes of different sensitivities. This figure-of-merit also assumes observations of single stars, and thus makes it difficult to compare to surveys targeting regions of the sky with a high density of stars, such as the center of the Milky Way or another galaxy. We did not attempt to use it.

\subsubsection{The Continuous Waveform Transmitter Rate}
The \BLI will carry out a variety of different surveys, from targeted surveys of nearby stars, to surveys of the Galactic plane and nearby galaxies \citep{2017PASP..129e4501I}. It would thus be beneficial to develop a figure-of-merit that allows us to more effectively take into account all the parameters of a search and compare the efficacy of a variety of different strategies.  Taking into account the limitations from other figures-of-merit outlined previously, we attempt here to create our own.

\begin{equation}
CWTFM = \zeta_{\rm{AO}}~ \frac{\rm{EIRP}}{ N_{stars}~\nu_{\rm{rel}}} \, ,
\label{eq:CWTFM}
\end{equation}
where $\nu_{\rm{rel}}$ is the fractional bandwidth $\Delta\nu_{\rm{tot}} / \nu_{\rm{mid}}$, with $\nu_{\rm{mid}}$ as the central frequency for a given survey. The total number of stars is defined as $N_{\rm{stars}} = n_{\rm{stars}}\times N_{\rm{pointings}}$, where $N_{\rm{pointings}}$ is the number of pointings during the survey, and $n_{\rm{stars}}$ is the number of stars per pointing. We assume $n_{\rm{stars}} = 1$ for targeted surveys. In future work, we will explore this assumption further to include stars in the background. We show the calculated values for this project and other SETI efforts in Tables 3 and 4. 
Finally, we define $\zeta_{\rm{AO}}$, as the normalization factor such that CWTFM =1 when EIRP = $L_{\rm{AO}}$, $\nu_{\rm{rel}}= 1/2$, and $N_{\rm{stars}} = 1000$. $L_{\rm{AO}}$ is the EIRP of the Arecibo Planetary Radar at $10^{13}$ W.

\begin{figure*}[htb]
\begin{center}
\includegraphics[width=0.95\linewidth]
{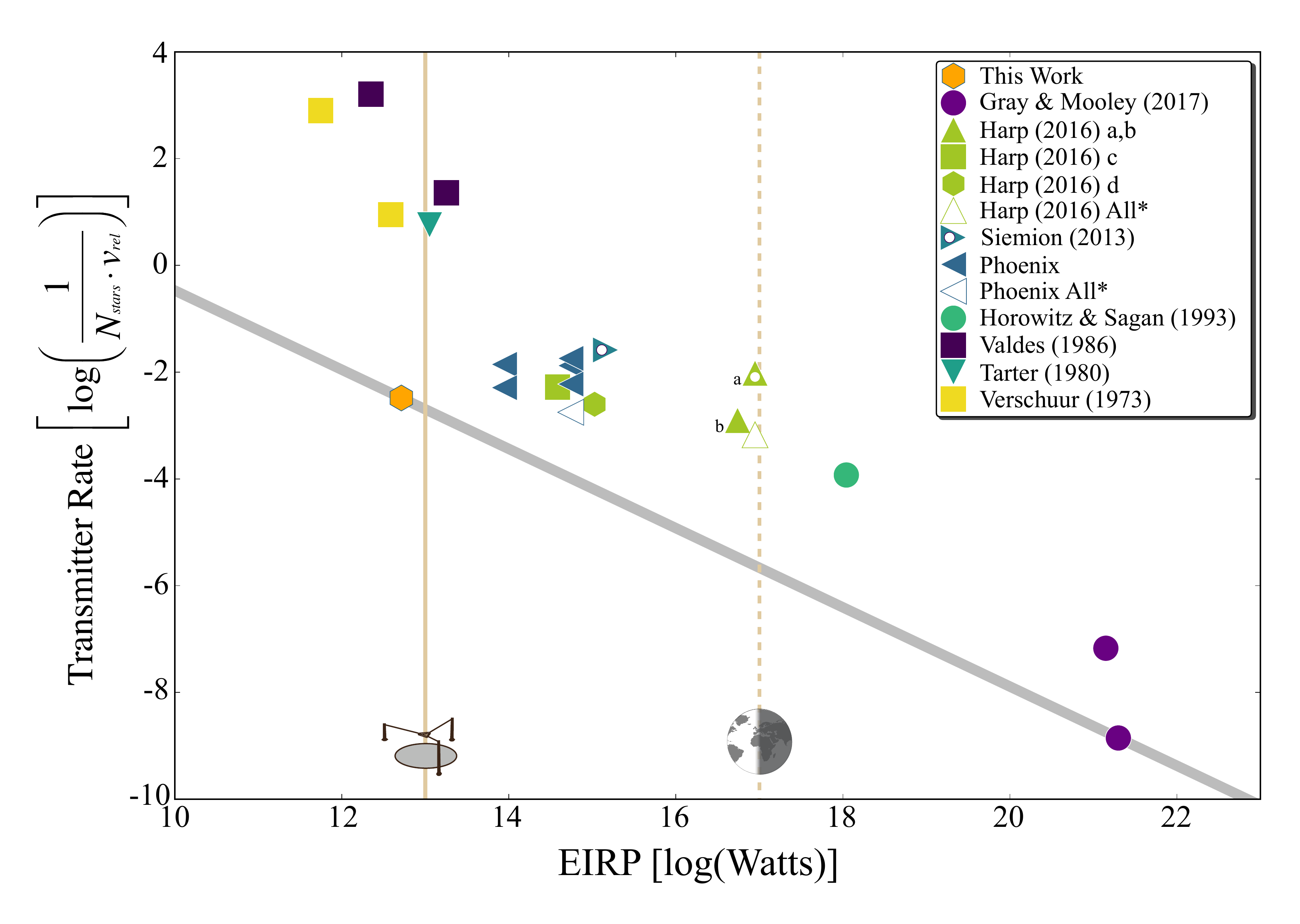}
\caption{Comparison of this work with several historic SETI projects. The vertical lines indicate characteristic EIRP powers, while the solid line indicates the EIRP of the AO planetary radar ($L_{\rm{AO}}$), and the dotted line indicates the total solar power incident on the Earth's surface, commonly referred as the energy usage of a Kardashev Type I civilization ($L_{\rm{KI}}$). The gray line is a fit of the values for this work and that of \cite{2017AJ....153..110G} by using Equation \ref{eq:power_law}. The points labeled "All", show the total for a given project, this value is calculated by the sum of Transmitter Rates and taking the largest EIRP.  
EIRP values were calculated based on the most distant target for a given survey;  sensitivity is better for nearer stars.
The total for other works with multiple surveys are not shown for clarity since they lie right on top of their lowest point. The shapes used for the different surveys is related to the stellar spectral types. Shapes with more sides indicate surveys targeting a wider array of spectral types. Triangles are used for searches only looking at solar-type stars (FGK) and circles are used to denote sky surveys with more than just main-sequence stars. The triangles with a white dot in the center show surveys targeting known exoplanets in the Habitable Zone. The values used to construct this figure can be found in Tables \ref{table:selected_searches} and \ref{table:selected_searches2}.}
\label{fig:galactic_rarity}
\end{center}
\end{figure*}

To visualize the previously compared surveys vis-a-vis the CWTFM, in Figure \ref{fig:galactic_rarity} we plot each survey's EIRP versus  $(N_{stars} \nu_{\rm{rel}})^{-1}$, we call the later the Transmitter Rate.

As shown in Figure \ref{fig:galactic_rarity}, this work provides the most stringent limit on low power radio transmitters around nearby stars, while the work from \cite{2017AJ....153..110G} does the same for the high power transmitters associated with nearby galaxies. This suggests that by using these two results together we can put a joint constraint on a luminosity function of artificial transmitters.

As has been done by others in the past \citep{1983sswg.book.....D,1985IAUS..112..411G,2000AcAau..46..649S}, we assume that the density of extraterrestrial transmitters in the galaxy follows a power-law distribution, which can be characterized as follows.
\begin{equation}
N(P_{tx}) = N_{0} ~ P_{tx}^{~-\alpha}  \, ,
\label{eq:power_law}
\end{equation}
where $N(P_{tx})$ is the number of transmitters as a function of power, $P_{tx}$. We assume an isotropic transmitter with $G_{ant} = 1$, and thus $P_{tx} = $EIRP.

Fitting between this work and \cite{2017AJ....153..110G} results in $\alpha \approx 0.74$ (indicated in Figure \ref{fig:galactic_rarity}), showing the transmitter occurrence space ruled out by this constraint. As a point of comparison, a fit to the EIRP of the strongest terrestrial radars shows a roughly power-law distribution with $\alpha \approx 0.5 $ \citep[][and references there in]{2000AcAau..46..649S}. 

We note here that as part of the \BLI, we plan to conduct a sensitive search of nearby galaxies with both Parkes and GBT. This search will be over a wide range of frequencies, improving constraints for very energetic transmitters.

We note that we used the most distant target to calculate EIRP sensitivity for most surveys we have compared to. However, detailed target lists were not always available and, in the case of the \textit{Kepler} field, in particular, distances are not well known.  For those cases, we used average or characteristic distance values. We favor the maximum distance since it is clear that all the stars in a given sample were observed to a given EIRP sensitivity. This statement is harder to maintain otherwise.
This approach has the issue of biasing the result toward the star with maximum distance, independent of the distance distribution of the group of stars. On the other hand, luminosity limited surveys would have the best scores, which may be a sensible result.  An obvious extension to this type of analysis would be to consider the entire distribution of stars within a radio telescope's primary beam, both near and far, when conducting an observation. 

\subsubsection{Other factors}

Despite the efforts here, many of the details of individual radio SETI experiments are difficult to capture in a single figure-of-merit.

One of the main aspects of an ETI search not taken into account in the figure-of-merit calculations presented here is the type of the search itself. As mentioned in Section \ref{sec:analysis}, the range and resolution of chirps searched provides an important extra parameter to sensitivity calculations. This is hard to quantify in many cases since it is, surprisingly, not always reported on the SETI literature. This differs markedly from the fast transient literature, in which the range of DMs searched is ubiquitously present. Also, it is difficult to quantitatively compare to some early work in which a correction to one or more ``special'' reference frames (e.g. Local Standard of Rest) is the only correction done.  In this burgeoning field, we encourage authors to clearly and fully describe all relevant aspects of their SETI searches.

Other aspects not included are mainly related to the potential anthropocentric biasing of a survey. For instance, most previous ETI searches look for intelligent life as we know it by looking only at solar-type stars. Nowadays, it is known that planets orbit stars of all spectral types.  We could then assume that intelligent life could live (if not evolve) around any star. 

We have also not treated additional selection constraints sometimes employed in SETI experiments, such as observations of stars with a transiting Earth-like planet orbiting in the Habitable Zone (HZ) or observations of stars in the Earth Transit Zone \citep[ETZ] {2016AsBio..16..259H}.

One last parameter not adequately covered in the analysis here is the frequency region observed.  As we move into an era where the exploration of wider frequency regions become possible, it will become increasingly important to consider the relative efficacy of observations well outside the $\sim$ 1$-$12 GHz terrestrial microwave window, at both lower and higher frequencies.

\section{Conclusions}
\label{sec:conclusion}

We have conducted a search for narrowband drifting signals toward 692 star systems selected from the original target list of the \BL\ project.
In an effort to reduce anthropocentric bias, we have searched stars across the full range of the main sequence.

Observations were performed with the L-band receiver on the GBT covering the range between 1.1 and 1.9 GHz.  The band was channelized into narrowband (3 Hz) channels, and a Doppler-drift search was performed to report hits consistent with a transmitter located outside of the topocentric frame of reference.
We determined that all the hits found by our algorithm are consistent with multiple types of anthropogenic RFI.

We find no evidence for 100\%-duty cycle transmitters (e.g. a radio beacon), either (1) directed at Earth with a power output equal to or greater than the brightest human-made transmitters, or (2) isotropic with a power output equal to the level of the current total human power usage on Earth, in any of the star systems observed.
Our results suggest that fewer than $\sim$ 0.1$\%$ of the stellar systems within 50\,pc possess these types of transmitters.

We explored several metrics to compare our results to previous SETI work. We note that the survey speed of the \BL backend is the fastest ever used for a SETI experiment by a factor of a few at least. Comparison with other SETI projects was also done by means of the DFM. We attempt to develop a new figure-of-merit that can encompass a wider set of parameters, to be used on future \BL experiments for a meaningful comparison.

The \BL\ project is ongoing, with new surveys planned, new detection algorithms being developed, and new telescopes brought online.
Beyond the classic narrowband search described in this paper, we are developing new methods to search voltage data, use data-driven model building for RFI classification, and image processing techniques to search for complex signals.

Over the longer term, the potential use of arrays such as MeerKAT \citep{2009IEEEP..97.1522J}, LOFAR \citep{2013A&A...556A...2V}, MWA \footnote{\url{http://www.mwatelescope.org/} }, ASKAP, \footnote{\url{https://www.atnf.csiro.au/projects/askap/index.html}} and others would provide an opportunity to search large numbers of stars ($\sim 10^{6}$) at a much faster survey speed compared to a single dish with equivalent sensitivity.  Furthermore, these facilities allow for commensal observations within a wide primary field of view to be conducted alongside other primary-user science observation programs.  These future surveys will provide increasingly strong statistical constraints on the space density of technologically advanced civilizations in the Milky Way, if not resulting in a detection of advanced extraterrestrial life. Observations of hundreds of galaxies could potentially provide estimates for the occurrence rate of the most advanced (Kardashev Type III ; \cite{1964SvA.....8..217K}) civilizations in the local universe.

\acknowledgments

Funding for \BL research is provided by the Breakthrough Prize Foundation\footnote{https://breakthroughprize.org/}.
We are grateful to the staff of the Green Bank Observatory for their help with installation and commissioning of the Breakthrough Listen backend instrument and extensive support during \BL observations.
We thank Jill Tarter and Gerry Harp for information provided on the Phoenix project, and Frank Drake for valuable comments.
We thank Angus Liang and Kevin Dorner for their generous support of undergraduate research at the Berkeley SETI Research Center. 
We thank Jason Wright for pointing out a typo in Equation 4.

\bibliographystyle{yahapj}
\bibliography{references}

\clearpage
\end{document}